\documentclass{article}
\usepackage[utf8]{inputenc}
\usepackage{graphicx}
\usepackage{natbib}
\usepackage{lscape}
\usepackage{hyperref}
\usepackage{color}
\usepackage{abstract}
\usepackage{ulem}
\newcommand{\footremember}[2]{%
    \footnote{#2}
    \newcounter{#1}
    \setcounter{#1}{\value{footnote}}%
}
\newcommand{\footrecall}[1]{%
    \footnotemark[\value{#1}]%
}

\usepackage{geometry}
\hypersetup{
    colorlinks=true, %set true if you want colored links
    linktoc=all,     %set to all if you want both sections and subsections linked
    linkcolor=blue,  %choose some color if you want links to stand out
}
\title{Anomaly Detection of Mobility Data with Applications to COVID-19 Situational Awareness}
\author{Stefano Maria Iacus\footnote{Corresponding author, Email: \href{mailto:stefano.iacus@ec.europa.eu}{stefano.iacus@ec.europa.eu} } \footremember{trailer}{European Commission, 
              Joint Research Centre,
              Via Enrico Fermi 2749, 21027 Ispra (VA), Italy} \and
        Francesco Sermi\footrecall{trailer} \and
        Spyridon Spyratos\footrecall{trailer} \and
        Dario Tarchi\footrecall{trailer} \and
        Michele Vespe\footrecall{trailer}
}

\date{}

\begin{document}

\maketitle

\begin{abstract}
This work introduces a live anomaly detection system for high frequency and high-dimensional data collected at regional scale such as Origin Destination Matrices of mobile positioning data. To take into account different granularity in time and space of the data coming from different sources, the system is designed to be simple, yet robust to the data diversity, with the aim of detecting abrupt increase of mobility towards specific regions as well as sudden drops of movements.
The methodology is designed to help policymakers or practitioners, and makes it possible to visualise anomalies as well as estimate the effect of COVID-19 related containment or lifting measures in terms of their impact on human mobility as well as spot potential new outbreaks related to large gatherings.
\end{abstract}
%\tableofcontents
\section{Introduction}
Mobile positioning data such as from Mobile Network Operators or social media, if received in almost real-time, have the potential to enhance situational awareness about events deviations from ``usual'' mobility patterns. Such anomalies may identify large gatherings that could be used as input to meta-population modelling and early warning applications aiming at flagging and projecting clusters that may lead to increases of $R_t$, the reproduction number.

Despite the fact that  mobility data alone cannot predict future needs, they can show already compelling citizens needs, like  transportation or heath care facility allocation needs and they represent well human behavior \citep{Bwambale20}. 
Moreover, thanks to the capability of collecting mobile data at very high time frequency and space granularity, the time evolution of mobility patterns can indeed show changes or ongoing trends or help to measure policy effects like the COVID-19 containment measures.

It is important to remark that, since mobile phone services unique subscribers\footnote{All  mobile services subscribers, including IoT, are about 86\% of the population, 76\% of which real smartphone users.} represent about 65\% 
of the population across Europe \citep{GSMA2020}, mobile data can reliably be used to capture the aggregate mobility patterns of the population.

In this work we present an anomaly detection system for mobile positioning data data able to handle data of different nature and formats, covering large areas and regions. Designed to process high volume and diverse input data, a robust system for anomaly detection was developed to detect not only excess of mobility but also missing or unexpected information in the data flow or sudden drops of mobility patterns.

This work is structured as follows. Section~\ref{sec:mobiledata} describes the input data in terms of volume, granularity and diversity. Section~\ref{sec:model} describes the basic idea of the anomaly detection systems and its scopes. Section~\ref{sec:limits} summarises the limits of the proposed approach.

\section{Mobile Positioning Data}\label{sec:mobiledata}

The system is designed to process data in the form of Origin-Destination Matrix (ODM) \citep{ODMs2019, ODMs2020,kishore2020}. Although the concept is somehow known to the general public, it is important to describe their nature to justify why the anomaly detection system of Section~\ref{sec:model} has to be designed simple yet robust to handle many different situations in a context of big and high frequency data.

%In order to deliver their telecommunication services, the MNOs need to collect information details like, e.g., the customer’s position, which needs to be constantly updated in order to route calls and data to the user. Two types of events are continuously being monitored: the Call Detail Records (CDR), which include mobile phone calls, messaging, and internet data accesses, and the eXtended Detail Records (XDR), which also include network signalling data. 

Each cell $[i-j]$ of the ODM shows the overall number of \textit{`movements'} (also referred to as `trips' or `visits') that have been recorded from the origin geographical reference area $i$ to the destination geographical reference area $j$ over the reference period.

To avoid any ex-post re-identification of individuals, before getting into an ODM, the data have to undergo several additional procedures such as deletion of any personal data, removal of singularities, thresholding, application of differential privacy (noise and distortions) methods and so forth.
In fact, ODM are usually shared in fully anonymised and aggregated form so that the risk of re-identification of individuals is virtually impossible.

\begin{figure}[!htb] 
    \centering
    \includegraphics[width=0.9\textwidth]{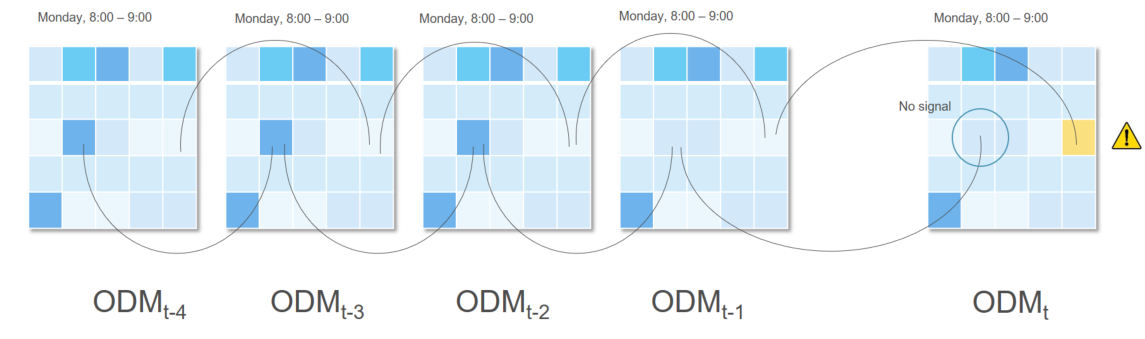}
    \caption{The simplified logic behind the anomaly detection strategy: a sudden drop of the volume of the cell may identify an anomaly, while one within the natural variability of the data not.}
    \label{fig:ODMs}
\end{figure}

In general, an ODM  (see also Figure~\ref{fig:ODMs}) contains the following  minimal information:
a timestamp for the \textit{start} and \textit{end} of the events considered, the areas of \textit{origin} and \textit{destination} and the \textit{counts} (movements, trips, etc).

In order to identify a movement between an origin area and a destination, it is necessary to define the \textit{dwelling} (or stop) time.
This dwelling time may vary from a few minutes to a few hours. A movement is recorded in the ODM only when the user stops for at least a duration equal to the dwelling time in the destination area having previously stopped for at least the same time in the origin area. An alternative way of defining a movement is to split the day in a number of time windows (normally 6- or 8-hour long) and to count the users that move from one geographical area to another between time windows; in this case, a user’s origin and destination areas are defined as those where the user spent most of the time in that time window. Also the definition of geographical area can be very different from one case to another: it can be an administrative area or a regular spatial grid. The construction of the ODM therefore depends on a number of tuning parameters. Depending on the choice of these tuning parameters, an ODM will be able to capture some types of movements but not others. For instance, an ODM may capture movements that extend for a long period of time but not shorter movements and vice-versa.

Despite the diversity of the ODMs that can be handled,  the ODM for a given source (social media or Mobile Network Operators) is consistent over time and relative changes are possible to be estimated. Some applications of these data to different contexts than the one presented here can be found in  \cite{SANTAMARIA2020104925,Iacus2020,TR-MFA}.

\section{A simple approach to anomaly detection}\label{sec:model}
%\section{Model Characteristics}
Detection of anomalies has a long history in statistics and quality control theory.  In the context of change point analysis for the location parameter one can see, e.g., \citet{bai97} and \citet{csorgo97} for  i.i.d. settings and \citet{bai94} for classical time series analysis, and in the context of  the scale parameter for several classes of processes, e.g., \citet{inclan94} and \citet{iacyos09}. These methods assume special data generation models and work with low dimensional and low frequency data mostly. In our case, we seek for robustness to data specification, computational efficiency and operational sustainability, therefore several decisions have been made to simplify the approach.

On one side, the anomaly detection system has:
\begin{itemize}
    \item to detect areas characterised by large increases of mobility that could be connected to gathering events; 
    \item to systematically provide data-driven knowledge of such events that can be input to real epidemiological early warning systems.
\end{itemize}
on the other hand the system has:
\begin{itemize}
    \item to detect sudden drops of data in input not only excesses as a system to detect data quality issues;
    \item to be computationally efficient given the dimensionality of the data in terms of frequency, spatial granularity and number of countries analysed;
    \item operationally feasible, i.e., produce almost real time and interpretable analysis;
    \item be robust with respect to high diversity of the input ODMs;
    \item be completely data driven in the sense that it should adapt itself to the time frame and granularity of the data.
\end{itemize}

To what extend the problem that the proposed system for anomaly detection has to address is related to handling high-dimensional data? As said, the ODM are generated by different sources with different time frequency and space granularity: the ODM can be as large as 10000 $\times$ 10000 entries time the 24 hourly sampling at country level. The system should be able to capture anomalies of two types: the excess of volume and the sudden drop of volume as well as unexpected filling of some elements of the sparse ODM matrix at hand. It has to consider a non symmetric approach, as sudden drops may well be related to error in input data, while and unexpected  excesses are structurally different and linked to large gatherings potentially critical in terms of Sars-Cov-2 spread. Being counts, the zero is a natural lower bound for low volumes times series, while the upper bound should be determined through standard statistical ideas.
We used a simple approach that takes into account both privacy thresholds (we not consider cells whose moving average is below the threshold $th$ (20 in our application), natural variability and moving average.
As it is well known that there exists both intra-daily, intra-weekly and seasonal patterns we apply short period moving average from the given date, time frame and space granularity.
Let $i$ be the origin, $j$ the destination, $s$ the start time and $e$ the end time of the sampling of the ODM for the date $d$. We denote each cell of the ODM matrix by
$$
ODM_{s,e}^d(i,j)
$$
where $i$ and $j$ spans the set of unique origin and destination labels $\mathcal C$, $d$ is a calendar date and $s$, $e$ are in the format $HH:MM:SS$.
If we want to consider the total inbound flow to a cell $j$, we use the notation 
$$ODM_{s,e}^d(\cdot,j) = \sum\limits_{i\in\mathcal C} ODM_{s,e}^d(i,j)$$ 
and we denote by 
$$ODM_{s,e}^d(i,\cdot) = \sum\limits_{j\in\mathcal C} ODM_{s,e}^d(i,j)$$ 
the outbound movements from cell $i$.
As there are situations in which the local movements are not interesting or such that the diagonal entries of the ODM matrix do not represent movements but people who stay in the same cell, we also consider the same quantities without the diagonals, i.e.,
$$\overline{ODM}_{s,e}^d(\cdot,j) = \sum\limits_{i\in\mathcal C, i\neq j} ODM_{s,e}^d(i,j)$$ 
and we denote by 
$$\overline{ODM}_{s,e}^d(i,\cdot) = \sum\limits_{j\in\mathcal C, j\neq i} ODM_{s,e}^d(i,j).$$ 
The moving average is take over the previous $p$ periods ($p=4$ in our aplication), i.e.,
$$
MA_{s,e}^d(i,j) = \frac{1}{p}\sum\limits_{t =1}^p ODM_{s,e}^{d-t}(i,j)
$$
and the rolling standard deviation is calculated similarly
$$
SD_{s,e}^d(i,j) = \sqrt{\frac{1}{p}\sum\limits_{t =1}^p \left(ODM_{s,e}^{d-t}(i,j)\right)^2 - \left(MA_{s,e}^d(i,j)\right)^2}
$$
In the event that for one or more past dates the data are not available, the $MA$ and $SD$ are calculated on the available data. If all past $p$ data are missing, no signal will be estimated and the date $d$ is marked as a ``missing data'' type. 
But historical variability in not enough as each ODM matrix, for different technical reasons at the MNOs level, may have a daily volume which is overall different from that of previous dates. This happens rarely but should be taken into account to avoid instrumentally false positives. Therefore, to take into account the overall variability, we select a first threshold $t$ corresponding to the 75\% quantile of the distribution of elements of the matrix   $ODM_{s,e}^d$ such that $ODM_{s,e}^d(i,j)\geq th$.
The upper bound is then set  to 
$$
U_{s,e}^d = \max(MA_{s,e}^d + t, MA_{s,e}^d + 3 SD_{s,e}^d),
$$
and the lower bound to
$$
L_{s,e}^d = \min(MA_{s,e}^d - t, MA_{s,e}^d - 3 SD_{s,e}^d, 0).
$$
So this is a simple 3-sigma  approach combined with a robust evaluation of daily variability.
More sophisticated time series approach or stochastic modelling (like inhomogeneous periodical Poisson process modelling) could have been used in spite of parametric tuning and estimation as well as computational time.
Indeed, the present approach has been chosen also because of the need of the speed of calculation. All the formulas above have been implemented in \texttt{R} \citep{rcore} via sparse matrix linear algebra and, whenever possible, calculation on the data base have been used to reduce the data transfer bottleneck.
The present approach can handle, for a single date, in less than an hour the analysis of several sources, providing data for up to 20 countries, at daily and, possibly, hourly frequencies. 
For example, for a single source, we have an ODM matrix of about 10000 $\times$ 10000 cells $\times$ and 25 time frames. The analysis is  performed also on the 10000 rows ($\overline{ODM}_{s,e}^d(\cdot,j)$) and 10000 columns separately ($\overline{ODM}_{s,e}^d(i,\cdot)$), considering the past 4 weeks as well (for the moving average calculation), i.e., the calculation of the anomalies is done on the non-null\footnote{Although many of the cells of the ODM matrix are empty being a sparse matrix, in a single day several thousands of them are not null and therefore should be considered in the analysis.} $(10000\times 10000 + 2\times 1000) = 100,020,000$ times series taking into account the 25 time frames for 5 dates (the present and the past $p$ dates). 
The signals are then marked as ``lower" and ``upper" signals and their intensity is evaluated in terms of relative increment with respect to the moving average. Let us denote this increment by
$$
INC_{s,e}^d(i,j) =\left(\frac{ODM_{s,e}^d(i,j)}{MA_{s,e}^d}-1\right) \cdot 100\%
$$
then, the level of the signal is classified as
\begin{itemize}
    \item level0 = no signal, i.e. $L_{s,e}^d(i,j)\leq DM_{s,e}^d(i,j) \leq U_{s,e}^d(i,j)$,
    \item level1 if $INC_{s,e}^d(i,j)< 50\%$, 
    \item level2 if $50\% \leq INC_{s,e}^d(i,j)< 100\%$, 
    \item level3 if $ INC_{s,e}^d(i,j)\geq 100\%$. 
\end{itemize}
for both lower ($ DM_{s,e}^d(i,j) < L_{s,e}^d(i,j)$) and upper ($DM_{s,e}^d(i,j)>U_{s,e}^d(i,j)$)  signals as well as for the inbound and outbound timeseries $\overline{ODM}_{s,e}^d(\cdot,j)$ and $\overline{ODM}_{s,e}^d(i,\cdot)$.
This type of filtering is helpful for the visual inspection of the thousands of signals appearing on a daily analysis. 

A possible extensions of this approach could consider also the spatial information contained in the data as in this approach the entries of the cells are considered independently (the only way they area considered together is using the overall quantile of the matrix). This type of approach will be computationally quite hard to solve and requires additional \textit{ad hoc} hypotheses according to the data source,  country and granularity, which we prefer not to use at this stage.

\newpage

\section{Conclusions and limits of this approach}\label{sec:limits}
As said, this simple and direct approach to the anomaly detection does not consider the spatial information contained in the data. This can be a nice addition in future developments of the system.
Indeed, parametric and non-parametric geo-statistical models can also be considered at the cost of putting assumptions on the data (by country and provider) and demanding for more computational time. The dimensionality of the problem is so high that, even using some restrictions like local dependency structure, it will become quite unfeasible to obtain model estimates in practical times though.

The system has been designed to alert on mobility anomalies for early warning capacity in case of COVID-19 outbreaks. 
Since these anomalies can be generally attributed to large gatherings and unusual mobility patterns in a broader sense, the system is a precious tool to understand the potential spread of the virus in case of outbreaks. At the same time, the system can allow authorities to monitor the implementation of mobility restrictions. 

The system is not designed to be a tracking system as it is totally agnostic to reality. It is also worth to mention that the system has not be designed to produce a real COVID-19 early warning system but only to spot anomalies in the data in the terms explained in Section~\ref{sec:model}. This means that there is no direct link in this application between, e.g., the large gatherings spotted and the reproduction rate $R_t$ of the COVID-19 pandemic.  Our data could only serve as an input to further epidemiological models or to policy makers to asses the effectiveness of the containment measures.

Despite its limitations, the systems seems to be able to capture what it is supposed to capture and  can accommodate different sources of ODM data without any stringent assumptions rather than the confidentiality threshold $th=20$, the length of the moving average $p=4$ and the quantile level $75\%$. These are the only three tuning parameters of the anomaly detection system and can be changed by the researcher.

% \paragraph{Funding statement}
%This research was supported by grants from the <funder-name><doi>(<award ID>); <funder-name><doi>(<award ID>).

\section*{Competing and/or conflict of interests}
None

\bibliography{references}

\bibliographystyle{chicago}

\end{document}